\documentclass[sigconf]{acmart} 
\pdfoutput=1
\usepackage[utf8]{inputenc}
\usepackage[T1]{fontenc}
\usepackage{xcolor}
\usepackage{url}
\usepackage{tabularx}
\usepackage{booktabs}
\usepackage{amsfonts}
\usepackage{amsmath}
\usepackage{pifont}
\usepackage{xspace}
\usepackage{tikz}
\usepackage{pgfplots}
\pgfplotsset{compat=1.14}
\usepackage{hyphenat}
\usepackage{balance}
\usepackage{nicefrac}

\definecolor{cycle1}{RGB}{228, 26, 28}
\definecolor{cycle2}{RGB}{55, 126, 184}
\definecolor{cycle3}{RGB}{77, 175, 74}
\definecolor{cycle4}{RGB}{152, 78, 163}
\definecolor{cycle5}{RGB}{255, 127, 0}
\definecolor{cycle6}{RGB}{153, 153, 153}\definecolor{cycle7}{RGB}{166, 86, 40}
\definecolor{cycle8}{RGB}{247, 129, 191}

\copyrightyear{2019}
\acmYear{2019}
\setcopyright{iw3c2w3}
\acmConference[WWW '19 Companion]{Companion Proceedings of the 2019 World Wide Web Conference}{May 13--17, 2019}{San Francisco, CA, USA}
\acmBooktitle{Companion Proceedings of the 2019 World Wide Web Conference (WWW '19 Companion), May 13--17, 2019, San Francisco, CA, USA}
\acmPrice{}
\acmDOI{10.1145/3308560.3316589}
\acmISBN{978-1-4503-6675-5/19/05}

\title{Spectral Graph Complexity}

\author{Anton Tsitsulin}
\affiliation{\institution{University of Bonn}
}
\author{Davide Mottin}
\affiliation{\institution{Aarhus University}
}
\author{Panagiotis Karras}
\affiliation{\institution{Aarhus University}
}
\author{Alex Bronstein}
\affiliation{\institution{Technion}
}
\author{Emmanuel M\"uller}
\affiliation{\institution{University of Bonn}
}

\begin{document}

\maketitle

\section{Introduction}\label{sec:introduction}

Graphs conveniently model relationships, interactions, physical and semantic structures.
On such graphs, data mining tasks, such as community detection and classification, are performed either directly or on low-dimensional embeddings~\cite{perozzi2014deepwalk,tsitsulin2018www}
derived from the graph structures.
By its nature, a graph represents complex interactions among objects in an abstract manner, while the underlying data intrinsically possess a notion of complexity.

Unfortunately, such complexity is hard to quantify  and assess, since the inherent shape of the data is unknown.
As such, there current way to have an implicit indication of the data complexity emerges after an analyst has engaged a graph in an involved data mining task.
Yet, the result of an analysis can be inexplicably bad or even meaningless, leaving no other choice than repeating the experiment without a precise indication whether said experiment will fail or succeed with different settings.
To avoid pointless analyses, we necessitate a characterization of the intrinsic complexity of the graph structure upfront.

Several works have tried to characterize some aspects of a graph's structure~\cite{faloutsos1999, broido2018} such as degree distribution or community structure.
Yet, none of them provides a characterization of the intrinsic complexity of general graphs.
Besides, recent efforts from the fields of chemistry and biology~\cite{rigterink2014, sinha2016} have been constrained to a domain-specific context, hindering the applicability to few specific cases~\cite{dehmer2009, sridhar2016}.

We take a different approach and study the graph from the lens of its spectrum. 
Spectral graph theory~\cite{chung1997} studies the properties of a graph based on the eigenvalues of the graph's Laplacian.
The full spectrum provides insights that allow for defining similarities between graphs~\cite{tsitsulin2018kdd}, and performing  various different tasks, such as spectral clustering~\cite{luxburg2007} or community detection~\cite{kloster2014}.
While the properties of the first non-zero eigenvalue are well understood especially in the context of algebraic connectivity~\cite{abreu2007, abbe2016}, the full characterization of the spectrum is still under investigation.
The Spectral properties of networks have been extensively used in quantum gravity studies, mainly in connection to synchronization properties~\cite{arenas2008}.
According to Weyl's Law~\cite{weyl1911}, the growth rate of the Laplacian eigenvalues is inversely proportional to the dimensionality of an underlying manifold.
The recently introduced complex network manifold model~\cite{bianconi2015} uses $d$-dimensional simplicial complexes as the building block for constructing more complex graphs with tunable spectral dimension.
Millan~et~al.~\cite{millan2018b} study synchronization properties of the complex network manifold model with relation to its spectral dimension.
However, a characterization of graphs in terms of spectral dimensionality has not been proposed before.

In this paper,
we introduce the \emph{first}, to our knowledge, universal complexity measure for general graphs which is directly deduced from the spectrum of graphs.
We show a preliminary result that demonstrates that the spectral dimension reflects the underlying complexity of real-world graphs by correlating spectral complexity measure to embedding quality on a large collection of real-world graphs.

\section{Spectral complexity estimation}\label{sec:solution}

 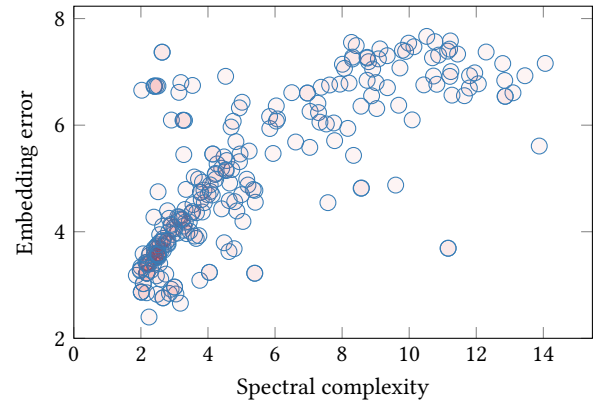
\begin{figure}[t]
	\begin{tikzpicture}
\begin{axis}[
		xlabel={Spectral complexity},
		ylabel={Embedding error},
		xmin=0,
ymin=2,
samples=150,
		height=6cm,
		width=\columnwidth,
]
\addplot+[only marks,mark=*,mark size=3pt,color=cycle2,mark options={fill=cycle1},fill opacity=0.05] table {data/chart-1k.data.txt};
\end{axis}
\end{tikzpicture}
\caption{Quality of VERSE embedding~\cite{tsitsulin2018www} correlates to our spectral complexity measure. Spearman correlation is 0.76, while mutual information equals 0.82.}\label{fig:verse-obj}
\vspace{-5mm}
\end{figure} 
We study the complexity of undirected graphs.
A graph is a pair $G=(V, E)$, where $V = (v_1, \ldots, v_n), n = |V|$ is the set of vertices and $E \subseteq (V \times V)$ the set of edges.
We assume the graph has no weights on edges even though our approach readily applies to the weighted case.

The \emph{adjacency matrix} of a graph $G$ is a $n\times n$ matrix $A$ having $A_{ij} \!=\! 1$ if $(i,j) \in E$ and $A_{ij}\! =\! 0$ otherwise.
A graph's \emph{normalized Laplacian} is the matrix $\mathcal{L}\! =\!  I\! -\! D^{-\frac{1}{2}}AD^{-\frac{1}{2}}$, where $D$ is the diagonal matrix with the degree of node $i$ in the  entry $D_{ii}$, i.e, $D_{ii} = \sum_{j = 1}^n A_{ij}$.
Since the Laplacian is a symmetric matrix, its eigenvectors $\phi_1, \ldots, \phi_n$, are real and orthogonal to each other.
Thus, it allows eigendecomposition as $\mathcal{L} = \Phi\Lambda\Phi^\top$, where $\Lambda$ is a diagonal matrix on the sorted eigenvalues $\lambda_1 \le \ldots \le \lambda_n$ of which $\phi_1, \ldots, \phi_n$ are the corresponding eigenvectors, and $\Phi$ is an orthogonal matrix obtained by stacking the eigenvectors in columns $\Phi = [\phi_1 \phi_2 \ldots \phi_n]$.

The set of eigenvalues $\{\lambda_1, \ldots, \lambda_n\}$ is called the \emph{spectrum} of a graph.
The normalized Laplacian, as opposed to the unnormalized version $L\! =\! D \!-\! A$, has a bounded spectrum, $0 \le \lambda_i \le 2$.
In general, the normalized Laplacian has more attractive theoretical properties than its unnormalized counterpart~\cite{vonluxburg2008}.

The normalized Laplacian matrix has the same spectrum as the random walk Laplacian $L^{\mathsf{rw}}=I-D^{-1}A$, which is closely related to the propagation of random walks in the graph. A random walk process starts from any node in the graph and at each steps randomly select one of the neighbor nodes.  
The random walk process can be studied in terms of its probability of occurrence of a random walker at node $i$ at time $t$ by the following equation:
$$
\pi_i(t) =  -\sum_j L^{\mathsf{rw}}_{ij}\pi_j(t-1).
$$
The \emph{spectral dimension} is commonly defined~\cite{durhuus2009} as the asymptotic behavior (at large times) of the probability of a random walk to return to the starting point $\pi_G(t)\sim t^{\nicefrac{-d_S}{2}}, t \rightarrow \infty$.
More formally,
$$
	d_S = -2 \lim_{t\rightarrow \infty}{\frac{\log \pi_G(t)}{\log t}}.
$$

However, this definition is useless for characterizing the complexity of finite graphs, as for every finite graph $d_S=0$ under this formulation, since the return probability is always $1$.
As real graphs are finite, we can leverage the fact that the spectral dimension can be estimated from the eigenvalue growth rate $\rho(\lambda)$~\cite{weyl1911}:

\begin{equation}\label{eq:weyls-law}
	\rho(\lambda) \simeq \lambda^{\nicefrac{d_S}{2}}
\end{equation}
for $\lambda\ll1$.
In $d$-dimensional Euclidean lattices $d_S=d$.
More generally, the spectral dimension relates to the Hausdorff (fractal) dimension which intuitively accounts for the local distances between points at multiple scales:

$$
d_H \geq d_S \geq 2 \frac{d_H}{d_H + 1}
$$
However, as far as we know, no result similar to the Weyl's law for manifolds has been obtained for finite graphs.
We take the analogy to the discrete case and build our spectral dimensionality estimator based on Weyl's law. 

Given a graph $G$ with $n$ vertices, 
we first compute its full spectrum or an approximation $\{ \lambda_k \}$.
We then linearly interpolate the spectrum producing a line $\lambda(x)$ within the interval $[0,1]$ such that $\lambda(k/n) = \lambda_k$.
The interpolated spectrum $\lambda(x)$ is then sampled on a fixed grid $(x_1,\dots,x_M)$ with $M$ points, producing an $M$-dimensional vector $\tilde{\lambda}$ with  entries $\tilde{\lambda}_k = \lambda(x_k)$.
The vector $\tilde{\lambda}$, having fixed size, is insensitive to the graph size and also invariant to the ordering of its vertices.
Finally, we select a point $s$ (in our experiments $s=\nicefrac{1}{100}$) and estimate the slope of the initial part of the spectrum $\tilde{\lambda} \leq \lambda(s)$. This spectrum slope approximates the asymptotic growth of the graph's eigenvalues.

\section{Preliminary experiment}\label{sec:experiment}

We downloaded 259 networks with size $10^3$ to $10^5$ from the Network Repository~\cite{rossi2015}, computed the spectrum of their normalized Laplacians, and estimated the spectral dimension as outlined in the Equation~\ref{eq:weyls-law}.
Next, we computed VERSE embeddings\cite{tsitsulin2018www}, using the default personalized PageRank (PPR) similarity, 128 dimensions, and $c=0.85$. We evaluate the capacity of VERSE of predicting the PPR similarity performance in terms of KL divergence with the real PPR vectors.

Figure~\ref{fig:verse-obj} shows strong relationship between our estimation of the spectral complexity and the objective performance on a graph task.
While there is some indication that not all networks are scale-free~\cite{broido2018}, we observe this phenomenon in terms of the spectral dimension, as only 3 out of 259 graphs have spectral dimension less than 2 which is related to scale-free networks.

As spectral dimensionality indicates the ``hardness'' of the graph for the embedding algorithms and embeddings are closely related to a variety of downstream tasks~\cite{tsitsulin2018www}, we envision important results in the understanding of data mining tasks on graphs. Our preliminary result attests the importance of the study of complexity measures for graph and defends the choice of spectral analysis as a privileged tool for the understanding of the performance of graph tasks. This study of infinite graphs provides rich insights for the finite case.

\balance
\bibliographystyle{ACM-Reference-Format}
\bibliography{bibliography} 

\end{document}